\documentclass{article}
\usepackage{spconf,amsmath,graphicx}
\usepackage{tabularx}
\usepackage{booktabs}
\usepackage{hyperref}

\title{NU-GAN: High resolution neural upsampling with GAN}
\name{Rithesh Kumar$^1$, Kundan Kumar$^{1,2}$, Vicki Anand$^{1}$, Yoshua Bengio$^{2}$, Aaron Courville$^2$\thanks{}}
\address{$^1$ Descript Inc. $^2$ Mila, Quebec Artificial Intelligence Institute}
\begin{document}
%
\maketitle
\begin{abstract}
In this paper, we propose NU-GAN, a new method for resampling audio from lower to higher sampling rates (\textit{upsampling}). Audio upsampling is an important problem since productionizing generative speech technology requires operating at high sampling rates. Such applications use audio at a resolution of 44.1 kHz or 48 kHz, whereas current speech synthesis methods are equipped to handle a maximum of 24 kHz resolution. NU-GAN takes a leap towards solving audio upsampling as a separate component in the text-to-speech (TTS) pipeline by leveraging techniques for audio generation using GANs. ABX preference tests indicate that our NU-GAN resampler is capable of resampling 22 kHz to 44.1 kHz audio that is distinguishable from original audio only $7.4\%$ higher than random chance for single speaker dataset, and $10.8\%$ higher than chance for multi-speaker dataset.
\end{abstract}
\section{Introduction}
\label{sec:intro}
Progress over the past few years in deep learning based techniques for speech generative modeling \cite{van2016wavenet, mehri2016samplernn, wang2017tacotron, prenger2019waveglow, kumar2019melgan} has allowed many practical applications of generative speech technology, such as text-to-speech (TTS) for powering voice assistants and characters in video games, audio denoising, speech separation, etc. These techniques have improved intonation and fidelity of speech synthesis systems \cite{van2016wavenet}. However, such methods are limited to generating a maximum audio resolution of 24 kHz. High-resolution audio (44.1 kHz) typically has much greater detail and texture due to its capacity to represent a wider range of frequencies, capturing sibilants and fricatives better to produce an overall crisper speech. Moreover, most streaming services and audio applications these days use the full high quality 44.1 kHz audio resolution. Therefore, to be useful in such applications, there is necessity for speech synthesis research to close this gap in audio quality.

\textbf{Difficulties in speech generative modeling:} Speech generative modeling is a difficult task since audio is inherently high dimensional (16,000 - 44,000 samples per second) with structure at different timescales, requiring the modeling of both short and long-term dependencies. Deep auto-regressive models are the first class of models that leverage deep learning to model raw audio successfully \cite{van2016wavenet, mehri2016samplernn}. However, these models suffer from slow sampling, often requiring multiple minutes to generate few seconds of audio. Innovation in learning algorithms such as GANs \cite{goodfellow2014generative} and Normalizing Flows \cite{rezende2015variational} permit fast training as well as parallel sampling. Recent works \cite{prenger2019waveglow, ping2018clarinet, kumar2019melgan, yamamoto2020parallel, engel2019gansynth} have explored these techniques for speech generation and successfully generate high quality audio with parallel, feed-forward sampling as well as training.

Despite their success, these methods are still limited to modeling audio at a resolution of 16-24 kHz and scaling to full resolution 44.1 kHz audio has not been successful yet. We hypothesize that this is because i) learning low frequency structure when modelling data at $2\times$ higher resolution is challenging since it requires modelling $2\times$ longer-term structure, ii) the inherent difference in the structure of audio in the low and high frequency ranges and how they correlate with the available conditioning information. Intelligibility of the speech depends on the low frequency components ($100 \text{Hz} - 11 \text{kHz}$) which are highly correlated with the high-level text information, whereas high frequency components ($11 \text{kHz} - 20 \text{kHz}$) determines the richness of the audio. These components are often uncorrelated with the high-level text, and are mostly dependent on the acoustic condition of the recording. Given poor correlation of these high frequency components with the available conditioning information e.g. text and speaker identity, learning a generative model for such structure is a challenging task. 

\textbf{TTS pipeline:} In this work, we primarily discuss parametric TTS aproaches, since they are the state of the art methods in terms of quality and intonation. TTS pipelines typically consist of 2 stages: 1) \textit{Text to spectrogram generation model}, 2) \textit{Neural vocoder} (spectrogram to waveform inversion model), although there are methods that directly attempt to generate raw waveform from text \cite{van2016wavenet, oord2017parallel}. In this paper, we propose an additional 3rd component in the TTS pipeline to perform neural upsampling, to convert lower resolution (16-24 kHz) audio to full high resolution (44.1 kHz) audio. This permits speech generation at full high quality without modifying the initial 2 stages in the pipeline. This is particularly beneficial since adapting the initial stages to support 44.1 kHz generation is not trivial.

\pagebreak
\textbf{Audio super-resolution / upsampling:} \textit{Audio super-resolution} is the task of converting audio from a lower to higher resolution. In the digital signal processing (DSP) literature, it is alternatively referred to as \textit{bandwidth extension}, \textit{sample-rate conversion} or \textit{resampling}, which is the process of changing the sampling rate of a discrete signal to obtain a new discrete representation of the same underlying continuous signal. Higher sample rates contain richer information since greater frequency ranges can be represented in the data. Most multimedia applications such as music and movies use audio at the highest resolution,  since they capture a greater level of detail and texture of various musical instruments. However, even speech applications benefit from high resolution audio, since they capture certain sounds in greater detail (such as sibilants and fricatives) apart from better modeling non-voiced sounds such as room tone.

\textbf{Contributions:} In this work, we propose NU-GAN, a novel high-resolution neural upsampling algorithm to generate high resolution audio at 44.1 kHz. Our method casts the problem of audio super-resolution into a waveform to waveform generation framework (within the same sampling rate). NU-GAN operates in the frequency domain rather than the waveform domain. This is beneficial for 2 reasons: 1) Only predicting the missing higher frequency bins in the magnitude spectrogram is an easier task compared to the task of generating raw waveforms, since the issues associated with phase estimation / generation can be sidestepped 2) Operating in the frequency domain is easier and compute-efficient compared to the waveform domain. Single-speaker and multi-speaker ABX tests indicate that our technique is almost indistinguishable from original 44.1 kHz audio.

\section{NU-GAN : Neural Upsampler GAN}
\begin{figure*}
  \includegraphics[width=\textwidth, scale=0.4]{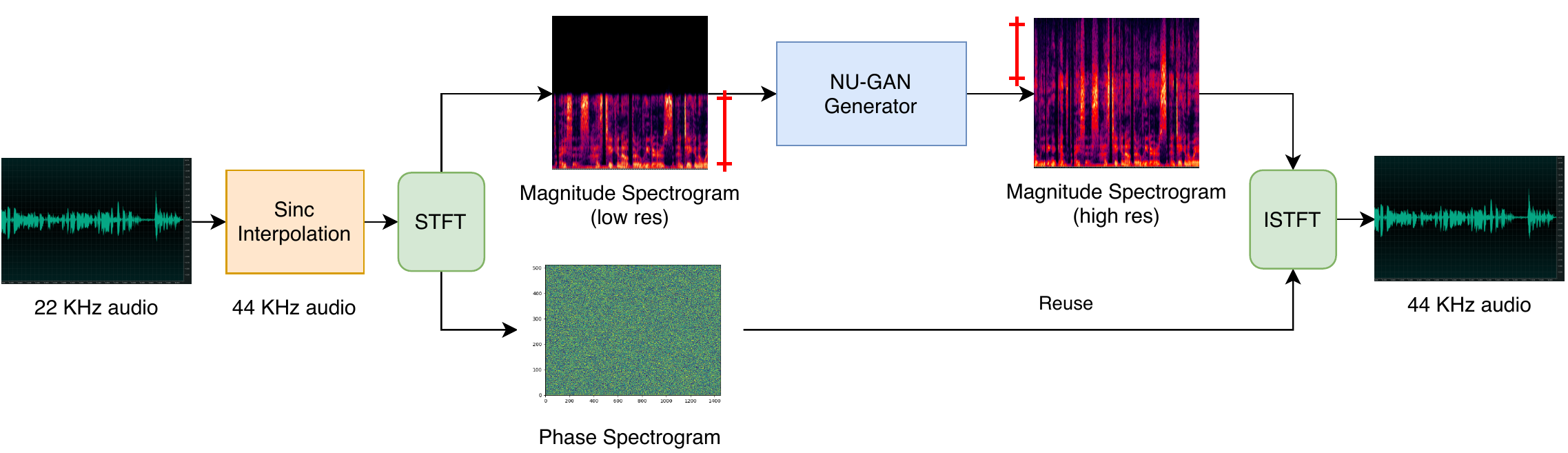}
  \caption{An illustration of our proposed technique to perform neural upsampling using NU-GAN}
  \label{fig:nugan}
\end{figure*}
\label{sec:format}
In this section, we describe the proposed technique to perform audio super-resolution. Figure \ref{fig:nugan} describes the overall pipeline. First, the low resolution audio (22 kHz) is upsampled to the full resolution (44 kHz) using the seminal Whittaker-Shannon interpolation formula (Sinc Interpolation) \cite{whittaker1915xviii}. Second, we compute the STFT using the interpolated audio and extract the magnitude and phase. Next, we train a conditional GAN model that predicts the missing higher frequency bins of the magnitude spectrogram conditioned on the lower frequency bins (pertaining to the lower resolution audio). Finally, the full resolution audio is generated by computing an inverse-STFT transformation with the full magnitude spectrogram and reusing the phase without alteration from the interpolated audio. This method of reusing the phase is sufficient for producing high quality waveforms, and is similar to speech enhancement and source separation techniques where the phase of the input audio mixture (mixture phase) is reused while synthesizing the individual sources.

We compute the STFT with 1024 filter, window length and 256 hop size, resulting in a magnitude spectrogram and phase with 513 frequency bins. Since we mostly deal with $2\times$ upsampling in our experiments, the input to the Generator network is the 257 lower frequency bins and the output is the top 256 higher frequency bins. While predicting the magnitude values, we use log scaling since human perception of sound approximates the logarithm of the intensity.

\subsection{Generator}
The task of the GAN generator is to predict the missing higher frequency ($256$) bins of the magnitude spectrogram (pertaining to true, high resolution audio at 44 kHz) conditioned on the lower frequency ($257$) bins. 

For this purpose, we use the popular transformer architecture. Our generator is a fairly simple model, consisting of a linear layer that transforms the 257 frequency bins of the input spectrogram to the model dimension of the transformer, followed by a series of 6 transformer layers. The output layer is a linear transformation that predicts 256 values, which correspond to the log magnitude values of the missing 256 higher frequency bins.

\subsection{Grouped Discriminators}
While the initial choice was to re-use the discriminator architecture from MelGAN \cite{kumar2019melgan}, we instead found it beneficial to alter the design to better suit frequency domain data such as magnitude spectrograms (rather than raw waveform).

Intuitively, we would like to independently process different frequency ranges in the magnitude spectrogram, learning discriminative features specific to each frequency range. While 1D convolutions can learn these features implicitly, we hypothesize that an explicit bias to process frequency ranges in groups is an efficient way to learn better discriminative features for our task, allowing us to learn a good generative model of all frequency ranges. While 2D convolutions can achieve this purpose, the translational invariance property is not well suited to learn features from spectrograms since patterns at low and high frequencies could be very different. In our work, we propose using 1D convolutions with groups \cite{xie2017aggregated} to bias our discriminators to learn features across different frequency ranges. Each layer in the discriminator groups the frequency bins into $n_\text{groups}$, where $n_\text{groups}$ is the same at each layer. We then use multiple discriminators, each of which group the frequency bins with different $n_\text{groups}$. This allows each discriminator to focus on different frequency bands in the generated output, while maintaining efficiency. The grouped discriminators therefore provide more direct feedback to smaller groups of frequency bins, helping the generator to better predict them.

In this work, we use 5 discriminators, with group sizes 1, 4, 16, 64 and 256, where the smallest  group discriminator treats the whole spectrogram frame as 1 group (coarse feedback), whereas the largest group discriminator treats each frequency bin in the spectrogram frame to be an individual group (fine feedback). Similar to MelGAN, we use a Markovian window-based discriminator architecture, with small modifications to better reflect the receptive field requirements while operating in the frequency domain. We found spectral normalization to be crucial, and is applied on all layers of the discriminator.



\subsection{Training Objective}
To train the generator and the grouped discriminators, we use the hinge loss version of the GAN objective \cite{lim2017geometric, miyato2018spectral}. Additionally we use the feature matching loss to train the generator \cite{kumar2019melgan}. Feature matching is applied on all but last layer of the discriminator, and on all 5 grouped discriminators.



\section{Experiments}
\label{sec:experiments}
In this section, we report quantitative metrics as well as qualitative metrics using our NU-GAN algorithm for audio upsampling from 22050 Hz to 44100 Hz. For this purpose, we test our algorithm on a single speaker internal dataset consisting of roughly 20 hours of audio of a female speaker recorded in a professional studio, as well as the \textit{cleanraw} subset of the publicly available DAPS dataset. DAPS dataset consists of 10 male speakers and 10 female speakers with each speaker reading 5 different scripts. We hold out 1 script (script5) for each speaker for testing and train on the rest of the data. We use the seminal sinc interpolation algorithm \cite{whittaker1915xviii} as our baseline in the experiments, which is also the input to our NU-GAN network. Hence, our results measure improvement in quality in comparison to the input interpolated 44100 Hz audio.

For all experiments, we use a 6-layer transformer Generator and an ensemble of 5 grouped discriminators (with group sizes 1, 4, 16, 64 and 256). Each discriminator is modeled similar to MelGAN \cite{kumar2019melgan}, except we use kernel size 4 stride 2 convolutions without any groups. We use the Adam optimizer to train the generator and discriminator with learning rates $0.0001$ and $0.0004$ respectively and betas $(0.0, 0.999)$.

\subsection{Quantitative Metrics}
In the signal processing literature, signal-to-noise ratio (SNR) and log-spectral distance (LSD) are popular metrics to measure audio reconstruction quality. While SNR compares the reference signal with the approximation in the waveform domain, LSD measures the reconstruction quality of individual frequencies. We found that our NU-GAN reconstructions produce poor SNR values since the output is not exactly time-aligned with the ground truth as our model is a GAN without an explicit reconstruction cost in the waveform domain. Therefore, we choose to report quantitative metrics in the frequency domain, since they do not harshly penalize small misalignments of the reference and approximate signals. The LSD metric is computed as follows:

\begin{align}
    \text{LSD}(x,y) = \frac{1}{L}\sum_{l=1}^L \sqrt{\frac{1}{K}\sum_{k=1}^K \big( X(l,k) - \hat{X}(l,k) \big)^2}
\end{align}

where $X$ and $\hat{X}$ are log-spectral power magnitudes of the reference and approximate signal respectively, defined as $X = \log |S|^2$, where $S$ corresponds to the STFT of the signal with filter and hann window of length 2048 and hop length of 512. 

Although it's hard to judge the effectiveness of the algorithm using quantitative metrics such as LSD, we report them here to help reproduce our results for future work. Scores are computed on a separate held-out set for both the internal single speeaker dataset as well as DAPS, as explained before.

\begin{table}[h]
  \caption{LSD scores on single speaker and multi-speaker datasets}
  \label{tab:benchmark-mos}
  \centering
  \small{
  \begin{tabular}{llc}
    \toprule
    Model     & Single-speaker  & Multi-speaker \\
    \midrule
    Baseline & $2.53 \pm 0.34$ &  $1.82 \pm 0.31$ \\
    \midrule
    \textbf{NU-GAN} & $\boldsymbol{1.43\pm0.10}$  &  $\boldsymbol{1.26\pm0.11}$ \\
    \bottomrule
  \end{tabular}
  }
\end{table}

\subsection{Qualitative Metrics}
To measure the perceptual quality of the results produced by the NU-GAN algorithm as compared to original 44100 Hz audio, we conduct listening tests using the ABX testing framework. Each testcase consists of 3 audio samples : A, B and X, where A and B correspond to the reference and approximate signals respectively and X is randomly sampled to be A or B. The rater is asked to listen to X first and find the sample A or B that is closest to B. The classification accuracy of raters is indicative of the distinguishability between A and B. $50\%$ accuracy (random chance) indicates that A and B are indistinguishable and $100\%$ accuracy on the other  hand indicates that A, B are identifiable all the time.

We conducted the ABX listening tests with sinc interpolation as well as NU-GAN as the candidate algorithms, compared to the ground truth 44100 Hz audio. To our best knowledge, this is the first work on audio upsampling that compares generated audio with ground truth high resolution audio, with competitive results. Each model was rated roughly 900 times using the amazon mechanical turk framework. The overall results of the ABX test are reported as follows.
\begin{table}[h]
  \caption{Average classification accuracy in the ABX test across all raters (lower is better).}
  \label{tab:benchmark-mos}
  \centering
  \begin{tabular}{llc}
  \\
    \toprule
    Model     & Single-speaker  & Multi-speaker \\
    \midrule
    Baseline & $77.5\%$ &  $71.4\%$ \\
    \midrule
    \textbf{NU-GAN} & $\boldsymbol{57.4\%}$  &  $\boldsymbol{60.8\%}$ \\
    \bottomrule
  \end{tabular}
\end{table}

The listening tests indicate that NU-GAN improves over the baseline sinc interpolation algorithm for a single speaker dataset by around $20\%$, and is identifiable only $7.4\%$ higher than random chance. In the multi-speaker dataset consisting of 20 speakers, NU-GAN improves over baseline by around $10\%$. 

\subsection{Text-to-speech synthesis}
To show the usefulness of our proposed NU-GAN upsampling algorithm in the TTS framework, we attach end-to-end TTS samples as part of the supplementary material as well as in our \href{https://ritheshkumar.com/nugan-icassp/}{\textbf{results website}}\footnote{Visit \href{https://ritheshkumar.com/nugan-icassp/}{https://ritheshkumar.com/nugan-icassp/} to listen to qualitative samples}. These samples serve as qualitative comparison between TTS at 22 kHz resolution vs 44 kHz resolution to showcase the importance of generating full resolution audio in modern neural TTS systems. Our TTS pipeline consists of: 1.) Text-to-melspectrogram generation component modeled using a network similar to FastSpeech but with added adversarial losses 2.) Mel-spectrogram inversion component modeled using MelGAN \cite{kumar2019melgan} and 3.) Neural upsampler modeled using NU-GAN. The overall pipeline converts text $\rightarrow$ mel-spectrogram $\rightarrow$ waveform at 22 kHz $\rightarrow$ waveform at 44 kHz.

\section{Related Work}
\label{sec:prior}
Audio super-resolution, also known as bandwidth extension was tackled in the seminal work of \cite{bansal2005bandwidth} using non-negative matrix factorization to perform upsampling of narrow-band (0-4 kHz) to broad-band (4-8 kHz) signals, with applications in speech telephony. \cite{mandel2015audio} improves over conventional NMF-based methods by performing unit selection in a concatenative synthesizer to generate clean recordings. More recently, deep neural networks (DNNs) have been applied to the same problem of narrow-band to broad-band conversion \cite{li2015deep}, outperforming traditional GMM-based approaches by directly mapping the missing high frequency spectrum. However, these methods are still limited to generating a maximum frequency resolution of 8 kHz, use hand-crafted features and relatively simple models. \cite{kuleshov2017audio} attempts to solve audio super-resolution by applying highly expressive image to image translation algorithms \cite{isola2017image} for audio translation, upsampling audio upto $8\times$ with a maximum resolution of 16 kHz. While \cite{kuleshov2017audio} casts audio super-resolution as an audio translation task in the waveform (time) domain, \cite{lim2018time} uses a time-frequency network (TFNet), processing the audio in both domains for better performance. These modern neural-network-based approaches fare better, but still lack the level of quality to be directly compared to ground truth high-resolution audio. Moreover, prior techniques have not yet attempted to generate audio at the full resolution of 44.1 kHz.

\section{Conclusion}
In this paper, we propose a novel approach to upsample audio. Specifically, we introduce a new stage in the TTS pipeline comprising of an audio-super-resolution module, which leads to broadcast quality generated speech. To our best knowledge, NU-GAN is the first to tackle audio super-resolution at a high sample rate of 44 kHz and is distinguishable from ground truth 44 kHz audio less than $10\%$ higher than random chance. NU-GAN is also fully-feedforward and fast to sample. We believe that our approach has applicability in non-speech domains as well, such as music. NU-GAN can also be utilized as a decompression algorithm in low-bandwidth applications that transmit audio at lower sampling rates.

\section{Acknowledgements}
The authors would like to thank Jose Sotelo and Prem Seetharaman for their support.


\vfill\pagebreak

\bibliographystyle{IEEEbib}
\bibliography{strings,refs}

\end{document}